\documentstyle{article}
\newtheorem{thm}{Theorem}[section]

\newtheorem{cor}[thm]{Corollary}

\begin{document}
\title{Using Mobile Agent Results to Create  
Hard-To-Detect Computer Viruses}
\author{Yongge Wang\\
Certicom Research, Certicom Corporation\\
5520 Explorer Dr., 4th floor, L4W 5L1, Canada\\
{\tt ywang@certicom.com}}
%\author{ }
\topmargin -0.7cm
\oddsidemargin 1.5cm
\date{ }
\maketitle
\abstract{
The theory of computer viruses has been studied by several
authors, though there is no systematic theoretical study 
up to now. The long time open question
in this area is as follows: Is it possible to 
design a signature-free (including dynamic signatures which 
we will define late) virus?
In this paper, we give an affirmative answer to this question
from a theoretical viewpoint.
We will introduce 
a new stronger concept: {\it dynamic signatures} of viruses,
and present a method to design viruses which are static signature-free
and  whose dynamic signatures are hard to determine 
unless some cryptographic assumption fails. We should remark
that our results are only for theoretical interest and may
be resource intensive in practice.
}

\section{Introduction}

The term {\it computer virus} is often used to 
indicate any software that can cause harm to systems
or networks. People often does not
include certain malicious software, such as Trojan horses and 
network worms, into the computer virus family.
In our discussion, a {\rm computer virus } refers to
any code that cause computer or network systems
to behave in a different manner than
the desired one.

A Trojan horse program is a useful or apparently
useful program or a shell script containing hidden
code that performs some unwanted function. A simple
example of a Trojan horse program might 
be a {\bf telnet} program. When a user invokes the program,
it appears to be performing telneting and nothing more,
however it may also be quietly changing
the file access permissions. Some Trojan horse
programs are difficult to detect, for example,
a compiler on a multi-user system that has been
modified to insert additional code into certain
programs when they are compiled (see Thompson \cite{thompson}).

The hidden code of a Trojan horse program
is deliberately placed by the program's author.
Generally, the hidden code in a computer virus 
program  is added by another program,
that program itself is a computer virus.
Thus one of the typical characteristics of 
a computer virus is to copy its hidden code to 
other programs, thereby {\it infecting} them.

Generally, a computer virus exhibits three
characteristics (see, e.g., \cite{pb,wca,wc}): 
a {\it replication} mechanism,
an {\it activation} mechanism, and an {\it objective}.
The replication mechanism performs the following
functions: search for other programs to infect,
when finds a program, {\it possibly} determines
whether the program has been previously infected, 
insert the hidden instructions somewhere in the program,
modifies the execution sequence of the program's instruction
such that the hidden code will be executed whenever the
program is invoked, and set some flags indicating that
the program has been infected. The flag may be necessary 
because without it, programs could be repeatedly 
infected and grow noticeably large.

The activation mechanism checks for the occurrence of some
event. When the event occurs, the computer virus
executes its objective, which is generally some unwanted,
harmful action.

Anti-virus tools performs three basic functions: 
{\it detect, identify,} or {\it remove viruses}.
Detection tools perform proactive detection,
active detection, or reactive detection. That is,
they detect a virus before it executes, during execution,
or after execution. Identification and removal 
tools are more straightforward in their application.

For detection of viruses, there are five classes of
techniques (see, e.g., \cite{pb}): 
signature scanning and algorithmic detection,
general purpose monitors, access control shells, checksums
for change detection, and heuristic binary analysis.
Note that after the paper  \cite{pb}, a new 
technique {\em emulation detection} has been
extensively used to detect polymorphic viruses.
In this paper, we will concentrate on
designing viruses which are signature-free and 
we will not discuss the techniques for identification
and removal.

A common class of anti-virus tools employs the
complementary techniques of signature scanning 
and algorithmic detection. This class of tools 
are known as scanners. Scanners are limited
intrinsically to the detection of known viruses.
In signature scanning an executable is searched for a 
selected binary code sequence, called a 
{\it virus signature}. 

General purpose monitors protect a system from 
the replication of viruses by actively intercepting
malicious actions. Access control shells function
as part of the operating system, much like
monitoring tools. Rather than monitoring 
for virus-like behavior, the shell
attempts to enforce an access control policy for the
system. Change detection works on the theory
that executables are static objects; therefore,
modification of an executable implies a possible virus infection. 
However, this theory has a basic flaw: some
executables are self-modifying. Heuristic
binary analysis is a method whereby the analyzer traces
through an executable looking for suspicious, virus-like behavior.
If a program appears to perform virus-like actions, a warning
is displayed.

Indeed, the signature scanner is the mostly used
technique to detect viruses. Note that at the beginning 
of this section we mentioned
that at the end of the replication mechanism, the virus
will generally put a flag in the infected program (this flag
can often be included in  the signature of the virus).
Otherwise the program may be repeatedly infected,
and the size may increase geometrically. Hence the virus will 
be detected easily.

The authors of computer viruses always try to design
viruses that are immune from those anti-virus software
on market. 
A {\it polymorphic virus} (see, e.g., \cite{pb,fs}) 
creates copies during replication
that are functionally equivalent but have distinctly different
byte streams. To achieve this, the virus may randomly
insert superfluous instructions, interchange the order of independent 
instructions, or choose from a number of different encryption schemes.
This variable quality makes the virus difficult to locate,
identify, and remove. This kind of virus can be 
thought as static signature-free viruses. However, up to my knowledge,
all known polymorphic viruses can be detected by algorithmic 
detection (or emulation detection) due to their dynamic signatures. 

In this paper, we will introduce a new concept: {\it dynamic signatures}
of viruses and present a method to design viruses
which are static signature-free and whose dynamic
signatures are hard to determine unless some cryptographic
assumption fails. This will answer a long time open question
in this area.

\section{Notations}
Turing machines are the basic model
for computation. We will present our
construction in terms of Turing machines.
Our notations are standard, e.g., as those in Hopcroft and Ullman \cite{hu}.
We will use two-way infinite tape, multi-track, and multi-tape Turing
machines as our model of computation. Formally, a Turing machine (TM)
is denoted 
$$M=(Q,\Sigma,\Gamma,\delta,q_0,B,F)$$
where $Q$ is the finite set of {\it states}, $\Gamma$ is the finite 
set of allowable {\it tape symbols}, $B$ is the {\it blank symbol}
which belongs to $\Gamma$, $\Sigma$ is the set of {\it input symbols}
which is a subset of $\Gamma$ not including $B$, 
$\delta$ is the {\it next move function} which is a mapping
from $Q\times \Gamma$ to $Q\times \Gamma\times\{L,R\}$ ($\delta$
may, however, be undefined for some arguments), $q_0\in Q$
is the {\it start state}, and $F\subseteq Q$ is the set of {\it final states}.
For each TM $M$, there is a read only input tape and a write only
output tape.

Without loss of generality, we will assume that $\Sigma = \{0,1\}$ 
in this paper unless specified otherwise.
The set of strings over $\Sigma$ will be denoted as $\Sigma^*$,
and the set of length $n$ strings will be denoted as $\Sigma^n$. 
A subset of $\Sigma^*$ is called a set, a problem, or just a language.
The characteristic function of a set $A$ is defined by letting
$A(x) = 1$ if $x\in A$ and $A(x)=0$ otherwise.

\section{Generating undetectable signatures}

From our analysis in previous sections,
in order for a virus to be undetectable, the virus must 
have some mechanism to check whether a program has 
already been infected by it or not.  For this purpose,
most viruses will put some flags in infected programs.
These flags generally can be considered as signatures.
Formally, the {\it signature} of a virus is a byte sequence
by which the virus can be distinguished from other programs
or viruses. That is, the virus code or programs
infected by it contain this byte sequence, but 
all other programs or viruses
generally do not contain this byte sequence.
This definition of virus signature 
is a static one (which is generally used in the literature). 
In this paper, we will introduce a new concept: 
a virus can have a {\it dynamic signature}.
Formally, a {\it dynamic signature} of a virus 
is a function which can be used to distinguish
it from other programs or viruses. From the definition,
it is clear that dynamic signatures of viruses
can only be determined by simulating the execution 
of the virus or the programs infected by it. 

It is clear that each non-trivial
virus should have either a static signature or a dynamic
signature. Polymorphic viruses do not have 
static signatures, but generally they have dynamic 
signatures. The reason why all previous polymorphic
viruses can be detected by 
current anti-virus software in market is that
their dynamic signatures are generally easy to determine.

Obviously, viruses which have static signatures
or whose dynamic signatures are easy to determine
can easily be detected. In order for a virus to be
undetectable, it should have the following properties:
\begin{itemize}
\item they have no static signatures in infected programs,
\item  their dynamic signatures are undetectable, 
for example, their dynamic signatures are hard to detect
unless some cryptographic assumption fails.
\end{itemize}
In this section, we present a method 
to design undetectable dynamic signatures.  
In the next section, we will present a method for
a virus to infect programs so that it leaves no static signature and in
the same time keeps the dynamic signature undetectable.

Let $F:N\rightarrow N$ be a secure function with the following
properties:
\begin{itemize}
\item $|F(x)| = |x|$ for all $x\in \Sigma^*$,
\item Given the values of $F$ on some 
given set $A\subset N$, it is hard for the enemy 
to compute any value of
$F$ on $x\notin A$,
\end{itemize}
where $N$ is the set of positive integers.
Such kind of functions can be chosen from any variant of digital
signature schemes or some encryption schemes, 
for example, a possible candidate is
the encryption scheme of  Cramer  and Shoup \cite{sc}
which is secure against 
adaptive chosen ciphertext attack. The reader is referred to 
\cite{mov} for more details on digital signature schemes
and encryption schemes..

A set $D$ is called {\it super sparse} if $|D\cap \Sigma^{n}|\le 1$
for all $n$ and $|D|=\infty$. For each virus we 
will assign a super sparse set $D$ as its dynamic signature.
Note that only the virus creator knows the dynamic signature 
of the virus.  It should also be the case that
it is computationally hard to guess the 
dynamic signature  of any virus.  
Let $n_0>100$ be a large enough integer and $F$ be a secure
function as mentioned above. Then
the dynamic signature for a virus is assigned as follows:
$$D_v=\{x\in\Sigma^n: x\mbox{ is the binary representation of }
F(n), n>n_0\}$$

Our method to insert the dynamic signature into a virus
is as follows: the virus will 
modify the behavior of the host program such that
the program will output random values of $0$ or $1$ on 
inputs from $D_v$.
Here we assume that the host program is a Turing 
machine $M$ which computes the characteristic 
function of a set $A$ (that is, $A$ means the normal behavior
of the program). Then the infected 
program will be a Turing machine $M^{\prime}$
with the properties that: $M^{\prime}(x)=A(x)$
for all $x\notin D_v$ and $M^{\prime}(x)$ is a
randomly chosen element from $\Sigma$ ($=\{0,1\}$)
for $x\in D_v$.
From the property of randomness, $M^{\prime}(x)= 1$
for approximately one half inputs $x$ from  $D_v$.

Note that if $n_0$ is large enough, then the set $D_v$
is a super sparse subset of $\Sigma^*$, whence the terminal
user will not notice the existence of the virus.
Also any anti-virus program will not be able to detect 
the virus unless one of the following conditions holds:
\begin{enumerate}
\item the anti-virus program can search a space of at least $2^{2n_0}$;
\item the function $F$ can be broken, that is, the anti-virus program 
can compute the value $F(n)$ for sufficiently many 
$n>n_0$. (Indeed, in practice,
the anti-virus program even does not know the 
function $F$).
\end{enumerate}

Let $M_{D_v}$ be the Turing machine  that
models a virus with a dynamic signature $D_v$.
That is, $M_{D_v}$ has the following
properties:
\begin{itemize}
\item $M_{D_v}$ computes the desired virus function (or behavior),
\item $M_{D_v}$ outputs randomly chosen elements
from $\Sigma$ for inputs from $D_v$.
\end{itemize}

\section{Checking infected programs}
\label{ccheck}

Each time when a virus find a chance to infect a program,
it will first decide whether it has already infected the
target program or not. For a virus with a static signature,
this will be straightforward. For a virus with a dynamic
signature (but without static signature),
this can be done by virtually simulating the 
execution of the target program on some inputs. 
The following algorithm can be used for this purpose.

\vskip 15pt
\noindent
{\bf Checking algorithm:}
\begin{enumerate}
\item Let $c=0, j=n_0+1$, and $K_0$ be a specific 
reasonably large integer (e.g., $K_0=100000$);
\item\label{loopbg} Let $x_j$ be the binary 
representation of $F(j)$. 
We distinguish the following two cases:

{\bf Case 1}: $M(x_j) = 1$.  Let $c=c+1$ and
$j=j+1$.

{\bf Case 2}: $M(x_j) =0$.  Let
$j=j+1$.

\item If $j < K_0$ then go to (\ref{loopbg}) else go to 
(\ref{looped}).
\item\label{looped} If $c\approx \frac{K_0-n_0}{2}$ then
$M$ has been infected. 
Otherwise, $M$ is not infected.
\end{enumerate}

\vskip 15pt

Note that the above checking algorithm sometime may 
output wrong information. That is, 
sometime it will consider a non-infected program as infected.
However, this will not be a big problem for a virus.
The purpose of a virus is to infect many programs,
and does not need to infect all programs.
Also it should be noted that if one can observe  one execution 
of the above checking algorithm, then one can easily
determine an initial part of the dynamic signature $D_v$,
whence help the anti-virus software to detect the virus.
In order to solve this problem, we can 
use the protocol of computation with encryption
(see \cite{af}).

Abadi and Feigenbaum \cite{af} described a protocol
how to securely evaluate a Boolean circuit in encrypted data.
They further reduced the problem of evaluation of
encrypted functions to the problem of processing
of encrypted data by representing the Boolean 
circuit that is to be hidden as data fed to a universal Boolean circuit.
Sander and Tschudin (\cite{st,st1})
have considered the similar problems in the context of
protecting mobile agents against malicious hosts.
They have shown that it is possible for a mobile agent
to actively protect itself against its execution
environment that tries to get some information of the
agent for some malicious goal. Especially, they have identified
a special class of functions -- polynomials and rational 
functions -- together with encryption schemes that lead to
a  non-trivial example of cryptographically hiding
a function such that it can nevertheless 
be executed with a non-interactive protocol. That is,
mobile agents executing this class of functions  can protect
themselves against tampering by a malicious host and can conceal the
programs they want to have executed. 
Obviously our above checking algorithm can be considered 
as a Boolean function of Abadi and Feigenbaum \cite{af}, though
generally it does not belong to the 
function classes identified by Sander and Tschudin (\cite{st,st1}).
In practice, a virus writer may find some trade-off 
between the results of Abadi and Feigenbaum \cite{af}
and Sander and Tschudin \cite{st,st1}, and 
achieve the following goal: it can check whether
a program has already been infected but not leak any information
of its dynamic signature.

From the discussion of this section, we can model
a virus with a dynamic signature $D_v$ as a Turing machine
$M_{D_v}$ with the following properties:
\begin{itemize}
\item $M_{D_v}$  computes the desired virus function (or behavior),
\item $M_{D_v}$ outputs randomly chosen elements
from $\Sigma$ for inputs from $D_v$.
\item $M_{D_v}$  can check whether a given program has already been
infected without leaking any information
about its dynamic signature  $D_v$.
\end{itemize}
In the next section, we will give a method to embed this
virus $M_{D_v}$ into any host program without leaving any
static signature.

\section{Inserting a virus into a program}

Assume that we have a Turing machine $M$ (that is, a program) 
which computes the characteristic function of a set $A$. 
That is, for each $x\in \Sigma^*$, $M(x) = A(x)$.
In order to infect $M$ with a virus $M_{D_v}$, 
we will combine the two Turing
machines into one Turing machine and 
introduce randomness into the codes so that
no static signature is left in the combined Turing machine.

For the reason of simplicity, we will assume that 
the virus Turing  machine $M_{D_v}$ 
will only compute the following function
$$M_{D_v}(x)=\left\{
\begin{array}{ll}
? & x\notin D_v\\
0\mbox{ or } 1 & x\in D_v
\end{array}
\right.$$
Of course, a virus will do many other harmful things in practice.
Our results can be easily extended to these harmful
viruses since we can regard the host program Turing
machine $M$ and the harmful part $M_v$ of the virus as 
one single Turing machine $\overline{M}$ and then apply our
method (note that it is trivial
to combine two Turing machines into one single Turing machine).
Now the following problem remains to be addressed:

\begin{itemize}
\item For each virus Turing machine $M_{D_v}$ and a host 
program Turing machine $M$, 
how to construct a Turing machine $M^\prime$
which computes the function
$$\mbox{vcomb}\{A(x), M_{D_v}(x)\}=\left\{
\begin{array}{ll}
A(x) & \mbox{if } M_{D_v}(x) = ?\\
M_{D_v}(x) & \mbox{otherwise}
\end{array}
\right.$$
such that
from $M^\prime$ it is computationally hard to get any information
of $D_v$? That is, how to construct $M^\prime$ such that
the anti-virus software cannot construct sufficient many
elements of $D_v$?
Note that the anti-virus software may have no time to run
the $M^\prime$s on all inputs to get some information
of the dynamic signature $D_v$.
\end{itemize}

\subsection*{One way vcomb-combination of Turing machines}

A Turing machine $M$ is called a {\it {\rm vcomb}-combination}
of TMs $M_1$ and $M_2$ if $M(x) =\mbox{vcomb}\{M_1(x), M_2(x)\}$
for all $x\in \Sigma^*$. A Turing machine $M$ is called a {\it composition}
of TMs $M_1$ and $M_2$ if $M(x) = M_2(M_1(x))$
for all $x\in \Sigma^*$.  
It is widely believed that in practice it is hard to 
decompose a Turing machine into two parts. For example,
a cryptosystem system based on the hardness of 
decomposition (and inverse) of finite automaton is suggested
in Tao and Chen \cite{tao}.
However, if $M$ is just a simple composition of
two TMs $M_1$ and $M_2$, then obviously one
can construct  $M_1$ (and $M_2$) from $M$ easily.

In the following, we will present a method for 
constructing the vcomb-combination $M$ of two TMs $M_1$ 
and $M_2$ with the properties that it is hard for the
adversary to construct a TM $M_2^\prime$
which is almost equivalent to $M_2$ in semantics,
that is, $M_2^\prime(x) = M_2(x)$ for sufficiently many (the exact
number depends on applications)
$x\in \Sigma^*$.

The procedure is as follows:

\vskip 15pt

Input: TMs $M_1$ and $M_2$ which compute $A(x)$ and
$M_{D_v}(x)$ respectively.

Without loss of generality, we may assume that 
$M_1=(Q_1,\Sigma,\Gamma,\delta_1,q_{1,0},B,F)$
and $M_2=(Q_2,\Sigma,\Gamma,\delta_2,q_{2,0},B,F)$
satisfy the following properties.

\begin{enumerate}
\item  $|Q_1|=|Q_2|$ and $|\delta_1|=|\delta_2|$;
\item  Both $M_1$ and $M_2$ are $3$ tapes Turing machines.
That is, $M_i$ has one read-only input tape, one working tape,
and one write-only output tape for $i=1,2$.
\item There is a number $k>0$ such that both $M_1$'s and $M_2$'s
working tapes have $k$ tracks\footnote{For any Turing machine $M$,
it is easy to change it to a $3$-tape (one input tape,
one $k$-track working tape, and one output tape) Turing machine.}.
\end{enumerate}

Now construct $M$ as follows:

\begin{enumerate}
\item Construct a 7-tape Turing machine $M_a$ with 
the following properties:
\begin{itemize}
\item The first tape is the input tape;
\item The second tape corresponds to the input tape of $M_1$;
\item The third tape corresponds to the $k$-track working tape of $M_1$;
\item The fourth tape corresponds to the output tape of $M_1$;
\item The fifth tape corresponds to the $k$-track working tape of $M_2$;
\item The sixth tape corresponds to the output tape of $M_2$;
\item The seventh tape is the output tape;
\item On input $x$, $M_a$ first copies $x$ from the first tape
to the second tape, then parallely simulates the 
computations of $M_1(x)$ and $M_2(x)$ (note that $M_2$ uses
the input from the second tape)
on the corresponding working tapes respectively.  In the end,
if the content on the tape $6$ is $?$, then 
$M_a$ copies the content on the tape $4$ to the tape $7$,
otherwise $M_a$ copies the content on the tape $6$ to the tape $7$.
And $M_a$ halts.
\end{itemize}
\item As in the proof of \cite[Theorem 7.2, pp.161]{hu}, convert $M_a$ into 
a 3-tape Turing machine $M_b$ with the following properties:
\begin{itemize}
\item The first tape is the input tape (the same as $M_a$'s input tape);
\item The second tape has $2k+8$ tracks, where the first two tracks
are for the second tape of $M_a$ (one to record the tape contents and one to
record the second tape's control head position), tracks $3$ through $k+3$ 
are for  the third tape of $M_a$ ($k$ tracks to record the tape contents
and one track to record the third tape's control head position), tracks $k+4$ 
through $k+5$ are for the fourth tape of $M_a$, tracks $k+6$ through
$2k+6$ are for the fifth tape of $M_a$, and tracks $2k+7$ through $2k+8$
are for the sixth tape of $M_a$.
\item The third tape is the output tape (the same as $M_a$'s output tape).
\end{itemize}
\item\label{onewayfunction} Choose a random permutation
$R:\Sigma^{2k+8}\rightarrow \Sigma^{2k+8}$.
Using the permutation $R$ to convert $M_b$ into a 3-tape Turing machine
$M$ with the following properties:
\begin{itemize}
\item  The second tape of $M$ has only one track. That is,  
the second tape of $M$ is divided into $k^{2k+8}$-size tuple-blocks.
Each tuple-block on the second tape of $M$ corresponds to one block of the
$M_b$'s second tape.
\item Each possible block on $M_b$'s second tape can be denoted
by a $2k+8$-tuple of symbols from $\Sigma$. The next move 
function $\delta_{M_b}$ of $M_b$ is converted to the next move function
of $\delta_M$ in such a way that each instantaneous description
block $x$ on $M_b$'s second tape (that is, a $2k+8$-tuple) is
mapped to an instantaneous description tuple-block $R(x)$
of $M$.
\end{itemize}
\end{enumerate}

\vskip 10pt
Note that generally it is the case that the anti-virus software
can get a copy of $M_1$ and $M$, and wants to find a description 
of $M_2$. In order for our construction to be robust against this
situation, we can make the following change to the vcomb-combination
construction: First make a random permutation of $M_1$ (similar
to what we have done for $M_b$ in the construction), then apply 
our above vcomb-combination construction. 
Now it is clear that if one can recover the $M_2$ from
$M$, then one can compute the permutation $R$.

\begin{thm}
\label{9870}
Let $f$ be an algorithm to construct $M_2$ from $M$, then $f$
can be used to compute the permutation $R$.
\end{thm}

{\bf Proof}. It is straightforward. \hfill$\Box$

\begin{cor}
\label{thisscor}
The probability that one can decompose $M$ into 
$M_1$ and $M_2$ equals to $2^{2k+8}!$.
\end{cor}

\begin{thm}
\label{thissthm}
There is an efficient process to construct from any two given 
Turing machines $M_1$  and $M_2$ a {\rm vcomb}-combination
Turing machine $M$ with the following properties: given $M$ and $M_1$,
with extremely high probability, one cannot get any information
of $M_2$.
\end{thm}

{\bf Proof}. This follows from Theorem \ref{9870}, Corollary \ref{thisscor},
and our above discussion. \hfill$\Box$

\subsection*{Signature undetectability}

Due to the random permutation $R$ in the vcomb-combination
process, it is straightforward that our virus will not have any 
static signature. The dynamic 
signature of our virus is also hard to detect.
First, by Corollary \ref{thisscor} and Theorem \ref{thissthm}, 
with extremely high probability
the anti-virus software cannot get a description of
Turing machine $M_{D_v}$ for the dynamic signature. Whence,
from a static analysis of the infected program, the anti-virus
software cannot get any information of the dynamic signature
of the virus. Secondly, when the anti-virus software simulates
a virtual execution of the virus (or the infected program),
it can monitor the process that the virus checks whether
a program has already been infected and the process 
that the virus inserts the hidden virus code into a program.
However, as mentioned in section \ref{ccheck}, special techniques
can be used to avoid the leakage of the dynamic sugnature.

\section{Conclusions}

Even though our results show that a virus could be written
in such a way that its dynamic signature is hard to detect,
each virus is still detectable. For example, when we notice 
the existence of a virus, we can write a simple program and
let it be infected. Then we can modify this infected program
as the virus detector as follows: run this program in a protected
environment and activate its virus infection code.
If this program decided not to infect the target code,
then with high probability that the taget code is 
already infected by this virus.
This implies that theoretically all known viruses can 
be detected. However this method is infeasible in practice.
The main difficulty here is that there may be millions 
of different viruses.
If we include all of these viruses into the anti-virus 
software package, the package will have a huge size,
and the detecting process will be extremely slow. 

In this paper, we have used some results from mobile agents
to improve the quality of viruses. Indeed, there is a 
close relationship between viruses and mobile agents.
In some sense, network worms could be considered as the 
prototype of mobile agents. Mobile agents have 
got extreme attention recently. We are sure that 
any future breakthrough in mobile agents protection will
also be a breakthrough for the design of undetectable
viruses.

As have been noticed by computer virus research community,
the best way to protect computer and network systems
against viruses is to use digital signatures. That is,
each time when a computer application package is 
developed, a digital signature of that software
should also be available to the users. This will
also defeat the viruses we have designed in this paper.
However, this is difficult to achieve in practice.
People like to download some shareware (e.g., games) 
from Internet and run it. For this kind of software,
you have to trust the author. Even if
you trust the share ware author, it is practically
difficult to include the signature keys of all shareware 
writers in the virus scanner. Hence the virus may be written
in such a way that it will only infect this kind
of shareware.

\end{document}